\documentclass[12pt,preprint]{aastex}

\shorttitle{Galaxy Bimodality}
\shortauthors{Forbes}

\def\etal{{\it et al.~}}

\begin{document}

\title{Bimodal Galaxies and Bimodality in Globular Cluster Systems}

\author{Duncan A. Forbes}
\affil{Centre for Astrophysics \& Supercomputing, Swinburne University,
  Hawthorn, VIC 3122, Australia}
\email{dforbes@swin.edu.au}

\begin{abstract}

Various galaxy properties are not continuous over a large range
in mass, but rather
reveal a remarkable transition or `bimodality' at a 
stellar mass of 3 $\times 10^{10}$
M$_{\odot}$. These properties include colors, 
stellar populations, Xray emission and mass-to-light
ratios. This behavior has been interpreted as the transition
from hot to 
cold flows by Dekel \& Birnboim (2005). 

Here we explore whether globular cluster (GC) systems 
also reveal a bimodal nature with regard to this critical
mass scale. Globular clusters probe star
formation at early epochs in the Universe and survive 
subsequent galaxy mergers and accretions.   
We use new data from the ACS Virgo Cluster Survey (Peng \etal
2005), which provides a homogeneous sample of the GC systems
around one hundred Virgo early-type galaxies covering a range of
five hundred in galaxy mass. Their classification of the GC color distributions is taken
to examine a key quantity -- the number of GCs per unit galaxy
luminosity. Below the critical mass, this quantity (called the
GC specific frequency) increases dramatically in its mean value and
spread. This increase may be due to regulated star formation in low
mass galaxies, which in turn is due to 
mass loss via winds and the transition from hot to cold gas
accretion flows. We also note that above the critical mass, galaxies possess two GC
subpopulations (with blue and red mean colors) but below this mass, galaxies reveal an increasing
proportion of single (blue) GC systems. 

\end{abstract}


\section{Introduction}

The general galaxy population reveals a bimodal color-magnitude
distribution (Lilly \etal 1995), which is seen in both the local
Universe by SDSS (Blanton \etal 2005) and out to modest redshifts
in COMBO17 (Bell \etal 2004). This blue/red bimodality shows a
well-defined transition at a stellar mass of about 
3 $\times 10^{10}$ M$_{\odot}$ or halo mass of 6 $\times 10^{11}$
M$_{\odot}$  (e.g. in SDSS by Kauffmann \etal
2003). Other galaxy properties, such as stellar populations (in
2dFGRS by Madgwick \etal 2003), Xray properties (O`Sullivan
\etal 2001) and galaxy mass-to-light ratios (Marinoni \& Hudson
2002) also reveal a bimodality about this critical stellar
mass (see Dekel \& Birnboim 2005 for a more complete list). 

This bimodality has been interpreted in terms of whether the
accreted material undergoes a virial shock when it enters the
galaxy halo (a {\it hot} flow) or not (a {\it cold} flow). Recent
simulations show that the transition from hot to cold mode
accretion occurs at the same critical stellar mass as the
observed galaxy bimodality (Keres \etal 2005; Dekel \& Birnboim 2005). 

It is of interest to know whether globular cluster (GC) systems
show a dichotomy along with the general galaxy population or if
their properties are continous over a large range in galaxy
mass. The spectra of GCs
in massive galaxies indicate that the bulk of the GCs are old,
i.e. $\ge$ 10 Gyrs with formation at epochs z$_f$ $\ge$ 2
(e.g. Strader \etal 2005a). 
The situation for low mass galaxies is
less well constrained. However, recent Keck spectra of 
GCs in the Virgo dwarf VCC 1087 indicate that they 
are also old and hence formed in the very early
stages of the dwarf galaxy formation (Beasley \etal 2005). Thus
GCs are formed during the earliest star and galaxy formation
processes in the Universe. Being of single age and chemical
composition, they do not suffer from the complication of multiple
star formation events as their host galaxies do.
They are also fairly robust, with many surviving the process of 
galaxy mergers and accretions. Globular clusters therefore
provide a useful probe of the early stages of galaxy formation
and their subsequent evolution.  

The number of globular clusters per unit
galaxy luminosity (called specific frequency) is 
a key measure of GC
systems, which varies with Hubble type and environment 
(see review by Elmegreen 1999). In this {\it Letter} we examine
how the GC specific frequency varies   
in relation to the critical mass and hence how it relates to 
the widely observed bimodality in galaxy
properties. 
From the Advanced Camera for Surveys (ACS) Virgo Cluster
Survey, Peng \etal (2005) noted that the only property of GC systems that
varied with the critical stellar mass was the fraction of red
GCs, i.e. it is nearly constant for high masses and declines
rapidly for
low masses. Strader \etal (2005b) studied the total specific
frequency for a subset of 37 Virgo dwarf galaxies (all with masses
below the critical mass). They found no strong difference between
dE and dE,N dwarfs, contrary to the earlier claims of Miller \etal
(1998) using WFPC2 data. Here, using the ACS data of Peng \etal, we
extend these earlier works  
to investigate the GC specific frequency for a homogeneous sample of
one hundred early-type galaxies in the Virgo cluster. 
We also briefly discuss the implications for GC and galaxy
formation.

\section{Observational Data}

The ACS Virgo Cluster Survey (Cote \etal 2004) consists of deep $g$
and $z$ band imaging of the 100 brightest early-type galaxies in
the Virgo cluster, using the ACS on-board the {\it 
Hubble Space Telescope}. This survey has the advantages of
increased depth, spatial coverage and metallicity sensitivity (to
detect any GC subpopulations) compared to previous WFPC2
studies (e.g. Forbes \etal 1996; Larsen \etal 2001; Kundu \&
Whitmore 2001). Blank fields were used 
to assist in determining the background contamination rate, which
is very low. The
integration times are such as to include $\sim90\%$ of the GC
luminosity function; the spatial coverage is complete for small
galaxies but severely incomplete for the GC systems of 
the largest galaxies (see below for further details). 

Recently Peng \etal (2005) examined the $g-z$ GC color
distributions for each galaxy in the survey (only VCC 1627 had no
detectable GCs). A number of galaxies revealed bimodal GC color
distributions in the largest homogeneous sample of GC systems to date. 
Such bimodality has been detected in the 
GC systems of numerous large galaxies (e.g. Zepf \& Ashman 1993;
Forbes \etal 1996; Larsen \etal 2001; Kundu \&
Whitmore 2001). This color bimodality indicates two distinct
GC subpopulations are present -- one blue and one
red. These subpopulations are thought, from spectroscopy, to be metal-poor and
metal-rich respectively, both being $\ge$ 10 Gyrs old with hints that
the red GCs may be slightly younger than the blue GCs
(e.g. Strader \etal 2005a).  
Other GC properties, such as their orbits and
spatial distributions, are also distinct when separated into
blue/red subpopulations. Peng \etal determined mean
colors for the blue and red subpopulations if the distribution
was statistically bimodal; for unimodal color distributions, the
GCs are all blue. They also list the 
number of GCs observed and the blue/red
fraction in bimodal systems. From this data we have determined
the specific frequencies of the blue and red GCs separately, i.e.\\

\noindent
S$_{Nblue} = f_{blue} \times N_{GC} \times 10^{0.4(M{_V} + 15)}$\hspace{3.9in}(1)\\
S$_{Nred} = f_{red} \times N_{GC} \times 10^{0.4(M{_V} + 15)}$\hspace{3.97in}(2)\\

\noindent
where $f$ is the fraction of blue or red GCs, N$_{GC}$ is the
total number of GCs and M$_V$ is the
galaxy absolute V-band magnitude calculated from the B-band
values listed in Cote \etal (2004) assuming B--V = 0.9 with no
reddening and m--M =
31.09. Ideally one would like to explore the relation between GC
subpopulations and the appropriate field stars within the galaxy,
but resolving individual stellar populations is currently only
possible for the very nearest elliptical galaxies (e.g. Harris,
Harris \& Poole 1999).

\section{Results}

In Fig. 1 we plot the S$_{Nblue}$ and S$_{Nred}$
separately against the host galaxy stellar mass.  To calculate
stellar masses we have assumed
a constant M/L$_V$ = 5 from the dynamical study of Virgo dwarf
galaxies by Geha \etal (2002). We note that a similar value is
obtained by Fukugita \etal (1998). This gives a range in mass
of a factor of $\sim$500.
The data points with an S$_{Nred}$ value of zero indicate the GC
systems that appear unimodal in color, i.e. the GCs are all
attributed to the blue subpopulation by Peng \etal (2005). 
For such systems we show typical errors.    
The plot also shows the critical stellar mass of 3 $\times
10^{10}$ M$_{\odot}$ (equivalent to M$_V$ = --19.6),  
about which the general galaxy population
reveals a bimodality in properties (see summary by Dekel \&
Birnboim 2005). The compact elliptical 
NGC 4486B (VCC 1297) has been excluded from this
plot as it has an abnormally high specific frequency of
S$_{Nblue}$ = 9.5 and S$_{Nred}$ = 4.1 (this is probably
due to a combination of tidal stripping reducing the galaxy's
luminosity and contamination from M87's extensive GC system; Peng
\etal 2005).

Figure 1 shows that below the critical mass galaxies reveal a
wide range in GC specific frequencies, whereas above the critical
mass the specific frequencies are confined to a relatively narrow
range. For masses below the critical mass we include a curve
showing a galaxy mass-to-light ratio M/L$_V$ $\propto$ M$^{-2/3}$ (Dekel \& Birnboim 2005), which
corresponds to L$_V$ $\propto$ M$^{5/3}$. The vertical
normalization is arbitrary. 

Below a mass of 10$^9$ M$_{\odot}$ the trend for higher
S$_N$ values continues (Durrell \etal 1996) but appears to halt at galaxy masses of $\sim$10$^{7}$
M$_{\odot}$ which have no GC systems (Forbes \etal 2000; Strader \etal 2005a). 
This lower mass limit for the occurrence of GC systems may
be set by atomic cooling, which is inefficient for virial
temperatures less than 10$^4$K corresponding to halo masses 
less than 10$^{8}$ M$_{\odot}$ (Moore \etal 2005).

Above the critical mass, both the red and blue specific frequencies are fairly
constant. For galaxies with masses 3 $\times$ 10$^{10}$ $<$
M/M$_{\odot}$ $<$ 3 $\times$ 10$^{11}$ the mean S$_{Nred}$ = 0.51
$\pm$ 0.05 and the mean S$_{Nblue}$ = 0.75 $\pm$ 0.06.
Figure 1 also shows the location of the bulge
specific frequency for M31 and the Milky Way from Forbes \etal
(2001), i.e. the total number of red GCs per unit bulge stellar mass.  
Within the errors, the S$_{Nred}$ of the bulges of these two spiral galaxies are
consistent with the trend seen in the Virgo early-type
galaxies. 

For the more massive galaxies, the ACS Virgo Cluster
Survey is missing a significant fraction of the GC system beyond its
3.4 $\times$ 3.4 arcmin field-of-view which would render the
S$_N$ value a significant underestimate, however this only
significantly affects a small
fraction of the overall survey. For example, the GC system of
Virgo's brightest galaxy M49 (VCC 1226) extends out to 23
arcmin and has a total GC population of 5,900 (Rhode \& Zepf
2001). 
For Virgo dwarf galaxies, Durrell \etal (1996) find that the
entire GC system is contained within 0.5 arcmin. The transition
from partial to full areal coverage by the ACS is not well
defined but occurs at intermediate galaxy luminosities. For example, a
wide area study of the GC system in the
low luminosity (M$_B$ = --19.35) Fornax elliptical NGC 1374
suggests about 83\% of the GC system would lie within the ACS
field-of-view if placed at the Virgo distance. This galaxy would
be the 16th brightest in the ACS Virgo sample. 
An additional, but smaller effect, for the brightest galaxies 
is that the blue-to-red GC ratio will also be
underestimated as blue GCs are preferentially located in galaxy
outer regions. 

\section{Discussion}

Figure 1 shows that GC specific frequencies reveal a large range
in values below the critical stellar mass, and that specific
frequencies are generally higher in lower mass galaxies. 
If GCs form at early epochs and with a similar
efficiency for all galaxies 
(McLaughin 1999; Moore \etal 2005), then some mechanism must be
reducing the stellar mass in low mass galaxies to raise the GC specific frequencies. 
Physical mechanisms for
regulating the stellar mass include: tidal stripping from the
interaction of a dwarf galaxy with a massive galaxy (e.g. Mayer
\etal 2001; Forbes \etal 2003), ram pressure stripping of dwarf
galaxies (van Zee, Skillman \& Haynes 2004), photoionization at early epochs
that suppresses further globular cluster and 
star formation (Beasley \etal 2002; Santos 2003; Bekki 2005;
Moore \etal 2005),  
supernova feedback and mass loss via winds (Durrell \etal 1996;
Dekel \& Silk 1986) and cold accretion flows (Keres \etal 2005;
Dekel \& Birnboim 2005). The latter has the advantage of
explaining the remarkable transition in GC specific frequency at the
critical stellar mass seen in Fig. 1, as well as the bimodality
in galaxian properties. 

We note that a halo 
mass-to-light ratio of the form M/L$_V$ $\propto$ M$^{-2/3}$
given by Dekel \& Birnboim (2005),  
equivalent to L$_V$ $\propto$ M$^{5/3}$, provides a reasonable
representation of the increasing GC specific frequency with
decreasing mass. However there is still substantial spread in the
S$_N$ values. In other words, below the critical
mass the number of GCs per unit galaxy mass appears to  
vary from galaxy to galaxy assuming the 
mass-to-light ratio varies in a systematic way with
halo mass. This may indicate a variation of baryonic to halo dark
mass.

Above the critical mass there is evidence for a fairly constant
S$_{Nred}$, i.e. a constant number of red GCs per unit galaxy
luminosity, which may extend to the bulge component of spirals as
first suggested by Forbes \etal (2001). This would support the
idea that spheroid growth is 
intimately linked to the formation of red GCs.

Interestingly, the correlation between the mean color of the GC
subpopulations and galaxy luminosity (or mass) does not show a
strong break or change in slope at the critical stellar mass. For
example, both Strader \etal (2005b) and Peng \etal (2005) found a
continuous GC color--galaxy luminosity trend in the ACS Virgo
Cluster Survey data used here. 
Thus the number of GCs per
unit galaxy luminosity and the blue-to-red fractions change at the
critical mass, whereas the individual properties responsible for GC
colors (e.g. age and metallicity) do not change abruptly. This is
consistent with the idea that GC individual properties are fixed at early
epochs, whereas galaxy properties (such as total luminosity)
can be modified by subsequent evolution. 


In terms of the evolutionary scenario outlined by Dekel \&
Birnboim (2005), low mass galaxies at high redshift 
are largely blue starforming disks which grow by accretion and cold
flows. Supernova feedback and mass loss regulate their star
formation. Shock-heating at a lower redshift in dense
environments shutoffs the cold flows, which in turn
suppresses any further star formation. These galaxies move
quickly onto the red sequence. Subsequent gas-poor (`dry')
merging allows the red sequence to extend from low mass galaxies to the most massive
ellipticals seen today. 

Star formation via cold accretion flows is thought to be very
efficient, thus to explain the higher specific frequency of low mass
galaxies as seen in Fig. 1 GC formation must be
even more efficient or the stellar luminosity is subsequently
reduced. In nearby galaxies for which we can assign the blue GCs (assumed to be the first
formed) to the appropriate fraction of the galaxy starlight (e.g. NGC 5128; Harris, Harris \& Poole
1999) high S$_{Nblue}$ values, and hence high GC formation
efficiencies, are inferred. So it may be possible that early GC
formation in {\it low} mass galaxies is similarly enhanced
relative to the star formation but this is very hard to verify
observationally. Reduction of the stellar luminosity (e.g. due to
supernova winds, stripping etc) is also a possibility.

An additional feature of the critical mass is that it represents
the onset of unimodal GC color distributions (with the exception
of the star-forming SB0 galaxy NGC 4340 (VCC 654) which may have
its luminosity enhanced by ongoing star formation). Thus
essentially all galaxies above the critical mass possess bimodal 
GC color distributions, i.e. evidence for blue {\it and} red GC
subpopulations. Below the critical mass, the frequency of
unimodal (i.e. blue only) GC color distributions increases, and
conversely the incidence of bimodality decreases, for
decreasing host galaxy mass (see Table 1). 

We speculate that this tendency towards blue-only GC systems may
be related to different evolutionary paths for dwarf
galaxies in cluster environments. For example 
if a bulgeless dIrr is transformed into a dSph/dE 
via tidal stripping (Mayer \etal 2001), then we might expect the
resulting dSph/dE to have few if any red GCs. 
The diversity of kinematic properties for dwarfs 
(both non-rotating and rotating systems have been found) 
suggests multiple evolutionary paths are possible (e.g. van Zee,
Skillman \& Haynes 2004).  
We note that the kinematic anisotropy parameter reveals
increased scatter for dwarf galaxies below the critical mass, 
reminiscent of Fig. 1.

\section{Conclusions}

Large surveys such as the 2dF, SDSS and COMBO17 have focused
attention on the fact that the general galaxy population reveals
a bimodality in properties at a critical stellar mass of $\sim 3 \times
10^{10}$ M$_{\odot}$.
These properties are largely bimodal in
nature, and include blue vs red colors, discrete vs diffuse
sources for the Xray emission and disk vs bulge domination.
To this list, we add another two properties
that reveal a bimodality around this critical mass, 
namely:\\

\noindent
$\bullet$ Above the critical mass galaxies reveal a
relatively narrow range in globular cluster specific frequencies,
whereas below they show an increasing mean value and spread.\\

\noindent
$\bullet$ Above the critical mass galaxies 
possess two globular cluster subpopulations,  below they
increasingly reveal single blue globular cluster populations. \\

Several physical mechanisms, which regulate star formation, may explain the increased specific
frequencies of low mass galaxies, however the observed transition
corresponds to the critical stellar mass which would therefore favor an
origin in supernova feedback and accretion flows as
suggested by Dekel and co-workers. 

\acknowledgements

We thank the ARC for financial support. We thank K. Bekki,
S. Larsen, R. Proctor, L. Spitler for useful discussions. 
The referee is also thanked for several useful suggestions.

\clearpage

\begin{figure*}
\includegraphics[scale=.65,angle=0]{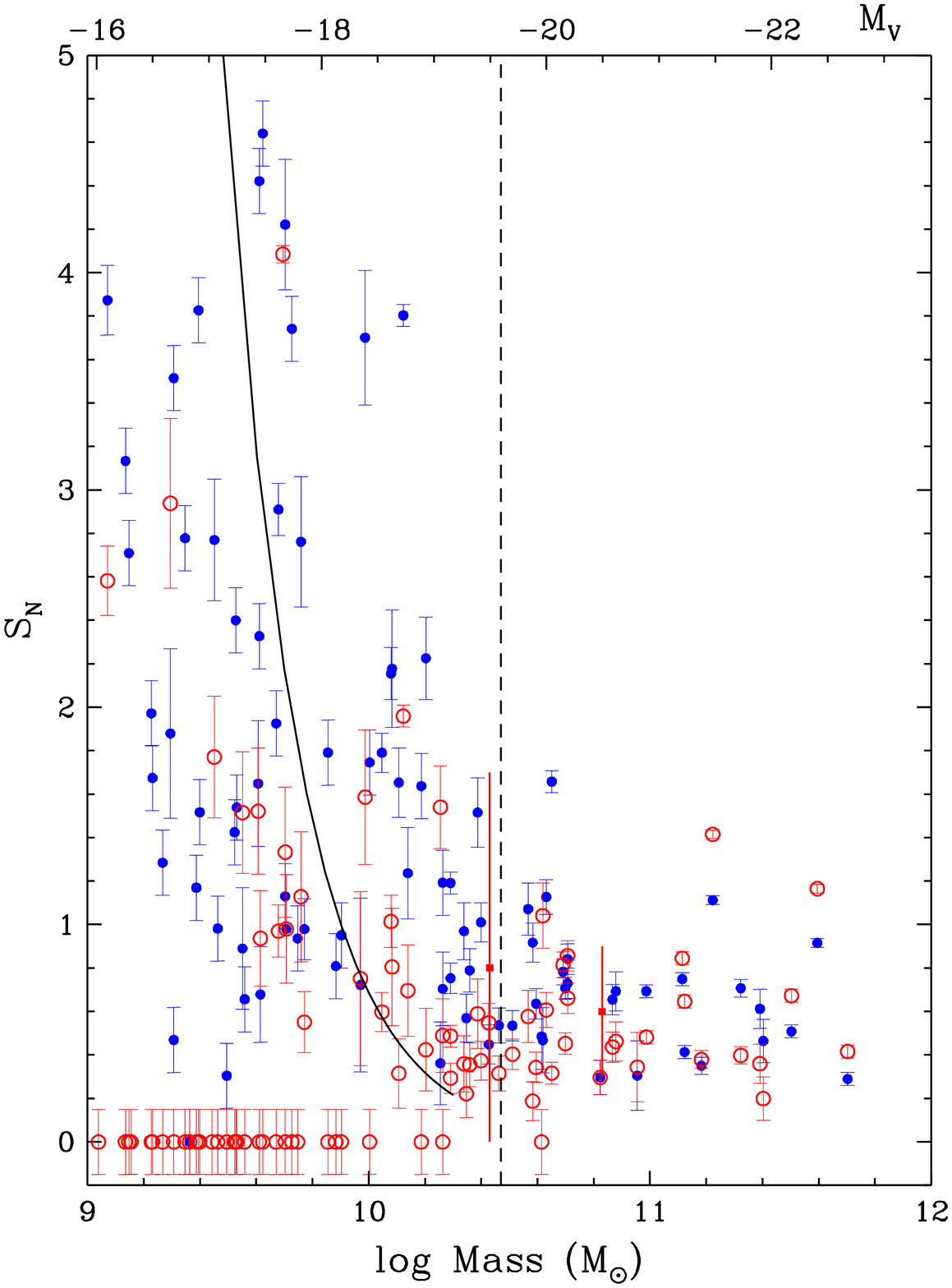}
\caption[fig1.ps]{
Globular cluster specific frequency S$_N$ versus host galaxy stellar mass. The blue
filled circles indicate the blue S$_N$ and the red open circles
the red S$_N$ values for Virgo cluster early-type
galaxies. Galaxies with no red GCs are indicated by red open
circles of value zero. The S$_N$ values of high mass galaxies are
underestimated due to the limited ACS field-of-view. 
A vertical dashed line is located at the critical stellar mass of 3
$\times$ 10$^{10}$ M$_{\odot}$ (or M$_V$ = --19.6, a constant
M/L$_V$ = 5 is assumed). Galaxies below the critical
stellar mass reveal a wide range of globular cluster specific
frequencies and an increasing proportion of galaxies have
unimodal blue GC systems. The red filled squares with error bars 
represent the bulge S$_N$ values for M31 and the Milky Way. The
solid curve represents a galaxy luminosity L$_V$ $\propto$ Mass$^{5/3}$ dependence
(see text for details).  
}
\end{figure*}

\begin{deluxetable}{lcc}
\tablewidth{0pc}
\tablecaption{Fraction of Unimodal GC Color Distributions}
\tablehead{Log Mass (M$_{\odot}$) & No. of Galaxies & \% Unimodal}
\startdata
$>$10.5 & 27 & 0\\
10.5--10.3 & 7 & 0\\
10.3--10.1 & 10 & 20\\
10.1--9.9 & 7 & 29\\
9.9--9.7 & 9 & 56\\
9.7--9.5 & 13 & 69\\
9.5--9.3 & 14 & 86\\
9.3--9.1 & 10 & 80\\
\enddata
\end{deluxetable}

\end{document}